\documentclass[%
mathleft,%
]{an}
\usepackage{graphicx}
\usepackage{times}
\begin{document}

\Pagespan{1}{}
\Yearpublication{2011}%
\Yearsubmission{2011}%
\Month{1}%
\Volume{999}%
\Issue{92}%

\title{Radio source evolution}

\author{M. Perucho\inst{1,2}\fnmsep\thanks{Corresponding author:
  \email{manel.perucho@uv.es}}
}
\titlerunning{Radio source evolution}
\authorrunning{M. Perucho}
\institute{
Departament d'Astronomia i Astrof\'{\i}sica, Universitat de Val\`encia, C/ Dr. Moliner, 50, 46100, Burjassot (Valencian Country, Spain)
\and 
Observatori Astron\`omic, Universitat de Val\`encia, C/ Catedr\`atic Jos\'e Beltran, 2, 46980, Paterna (Valencian Country, Spain) }

\received{XXXX}
\accepted{XXXX}
\publonline{XXXX}

\keywords{galaxies: active, galaxies: jets, hydrodynamics.}

\abstract{%
 Baldwin (1982) wrote that \emph{the distribution of sources in the radio luminosity, $P$, overall physical size, $D$, diagram} could be considered as \emph{the radio astronomer's $H-R$ diagram}. However, unlike the case of stars, not only the intrinsic properties of the jets, but also those of the host galaxy and the intergalactic medium are relevant to explain the evolutionary tracks of radio radio sources. In this contribution I review the current status of our understanding of the evolution of radio sources from a theoretical and numerical perspective, using the $P-D$ diagram as a framework. An excess of compact (linear size $\leq 10$~kpc) sources could be explained by low-power jets being decelerated within the host galaxy, as shown by recent numerical simulations. Finally, I discuss the possible tracks that radio sources may follow within this diagram, and the physical processes that can explain the different tracks.}
  
\maketitle

\section{Introduction}
 
   Baldwin (1982) suggested that radio-galaxies follow evolutionary tracks within the $P-D$ diagram, in which the sources are represented in terms of their radio luminosity $P$ and their linear size $D$ (see, e.g., Fig.~1 in Baldwin 1982, Fig.~1 in An \& Baan 2012). At the time, selection effects of the available surveys were hiding the low-lu\-mi\-no\-si\-ty population of radio sources. More recent surveys of low-luminosity sources (e.g., de Vries et al. 2009, Kunert-Bajraszewska et al. 2010) have revealed a number of sources distributed throughout the diagram. A  classification adapted to the new scenario revealed by low-lu\-mi\-no\-si\-ty observations has been proposed by An \& Baan (2012), including predicted tracks depending on the initial power of the source and the activity period of the active galactic nucleus (AGN). However, a dynamical explanation for different tracks followed is required. Namely, FRI (typically low-power and showing decollimated and disrupted structure at kiloparsec scales, Fanaroff \& Riley 1974) and FRII (typically high-power and showing collimated structure at kiloparsec scales) radio-sour\-ces are expected to follow different paths through this diagram. Moreover, a new class of objects has been recently proposed on the basis of their lack of structure at large scales (Baldi et al. 2015), and have been named FR0 accordingly. In this contribution I discuss not only the role of the intrinsic jet properties, but also that of the ambient medium in the long-term evolution of jets.
   
   A number of theoretical and numerical models have been developed to explain the evolution of radio sources within the inner kiloparsec (e.g., De Young 1993, Fanti et al. 1995, Readhead et al. 1996, Perucho \& Mart\'{\i} 2002) and beyond (e.g., Begelman \& Cioffi 1989, Kaiser \& Alexander 1997, Perucho \& Mart\'{\i} 2007, Perucho et al. 2011, Maciel \& Alexander 2014). These models are successful to explain the overall evolution of powerful radio sources to first order. However, the detailed dynamics require numerical simulations to be fully understood. In this respect, recent numerical simulations of jet evolution through a clumpy medium within the first kiloparsec of evolution (see, e.g., Wagner et al. 2012 for the case of a relativistic outflow, and Gaibler et al. 2012 for a Newtonian one)  have shown that the propagation of the jet can be extremely complex in this region due to interaction with large clouds and that the forward shock could trigger star formation under the appropriate conditions. In addition, relativistic hydrodynamic (RHD) simulations of long term evolution of jets beyond the first kiloparsec in a decreasing density ambient medium show that analytical models give a correct overall picture of the evolution of the cocoon/cavity and shocked region (e.g., Perucho \& Mart\'{\i} 2007, Perucho et al. 2011). The simulations also reveal that the cocoon is surrounded by a region of shocked ambient medium and the whole structure is surrounded by an external (bow-)shock (Perucho et al. 2011, 2014b). These shocks are elusive to observations and only during the last decade they have started to be detected via modeling of X-ray observations (e.g., Nulsen et al. 2005, McNamara et al. 2005, Simionescu et al. 2009, Gitti et al. 2010, Croston et al. 2011, Stawarz et al. 2014). 
      
   It is relevant to make a note on the vocabulary related with the evolution of radio sources from the point of view of numerical simulations and observations. In the former, we refer to the \emph{backflow} as the shocked jet gas that propagates in the opposite direction to the jet flow. This region is also referred to as \emph{cocoon}. Radio observations reveal the regions of the backflow closest to the hotspots in FRII jets, which are named \emph{lobes}. The cocoon is, in simulations, the origin of X-ray \emph{cavities}. When I refer to the size of the observed radio sources, I will be referring to the linear size of the radio-struc\-tu\-re that is associated to the jet. An alternative would be to define it as the linear size of the bow-shock. However, bow-shocks are actually not seen in radio, although a hint is given by the shape of the radio-emitting lobes in powerful radio sour\-ces. Therefore, it is difficult to find traces of them for low-power sources. 

    This contribution is structured as follows: In Section~\ref{dm} I present different possible deceleration mechanisms that can slow or stop the growth of the radio source, Section~\ref{et} is devoted to the different evolutionary tracks in the $P-D$ diagram, and in Section~\ref{di} I discuss critically the results listed in the previous sections and review the open issues in the field.

\section{Deceleration mechanisms}\label{dm}
 
   My main aim in this manuscript is to give a description of the evolution of radio sources in terms of their size and their luminosity. As far as the former, it is obviously related to the advance velocity of the jet reverse shock. This has to be differentiated from the advance velocity of the bow-shock that is triggered in the ambient medium by the injection of the supersonic jet: The jet reverse shock (followed by the hot-spot in collimated radio jets), in observational terms, is attached to the bow-shock in collimated jets, but this is not the case for those flows that are disrupted or decelerated to trans-sonic velocities. In this case, the head of the jet is detached from the bow-shock, which propagates driven by the pressure difference between the cocoon/shocked ambient region and the unshocked ambient medium. 
    
   The advance velocity of a relativistic jet (in a one-dimen\-sio\-nal calculation) is given by (Mart\'{\i} et al. 1997): 
\begin{equation}
\displaystyle{ v_{a}\,=\,\frac{\sqrt{\eta^*}}{1+\sqrt{\eta^*}}\,v_j },
\end{equation} 
\noindent where $v_j$ is the velocity of the jet flow and $\eta^*\,=\,\eta \, \gamma^2$, with $\gamma$ the jet Lorentz factor and $\eta$:
\begin{equation}
\displaystyle{\eta\,=\,\frac{\rho_j\,h_j}{\rho_a\,h_a}},
\end{equation} 
where $h$ the specific enthalpy and $\rho$ the rest mass density, and subscripts $j$ and $a$ stand for the jet and the ambient medium, respectively. All the quantities are given in the reference frame of the host galaxy. 

      This expression clearly states that the larger $\eta^*$, the closer is the advance velocity of the head to the jet flow velocity, and that $v_a$  velocity is proportional to $\gamma v_j$. Taking into account that $h_a \simeq 1$ (in units of $c^2$) in typical interstellar and intergalactic media (ISM and IGM, respectively), one needs extremely dense or hot jets to obtain large values of $\eta^*$. In the case of relativistic jets, the presence of lobes indicates that $v_j$ has to be larger than $v_a$ ($v_a \rightarrow v_j$ only when $\eta^* \gg 1$), so we still expect $\eta^*\leq 1$ in most cases. Changes in the value of $\eta^*$ should not have an important effect on $v_a$ because: 1) An increase of the density of the jet flow (that would produce a corresponding increase in $v_a$) due to entrainment would also lead to deceleration of the jet (a decrease in $v_j$) due to the conservation of momentum; 2) an increase of the temperature could occur via dissipation of kinetic or magnetic energy into particle internal energy at shocks or interactions (increasing $\eta^*$ and $v_a$ accordingly), but this process would also lead to jet flow deceleration (decreasing $v_j$ and $v_a$ accordingly) . Therefore we can focus on the different processes that can change the parameter $v_j$ (and therefore $\gamma$) because it is the most relevant to determine $v_a$ from the one-dimensional approximation, which can give us hints on the evolution of radio sources in size. In the next subsections I discuss the effects of the growth of instabilities, of mass-load by stars and clouds embedded in jets, and shocks on jet velocity.      
     
\subsection{Instabilities}
    
     Relativistic jets are flows of particles and energy, and are thus subject to the growth of Kelvin-Helmholtz (KH) and/or current-driven (CD) instabilities. The former grow at the contact discontinuity (or shear-layer) between two flows with relative velocity tangential to the discontinuity, whereas the latter grow in strongly magnetised flows, when the field lines are distorted (see Hardee 2011, Perucho 2012 for recent reviews).  Although jets are formed in strongly magnetised regions around the central black holes in active galaxies, it is still a matter of debate whether jets are magnetically dominated or particle dominated at parsec-scales and beyond (see, e.g., Nakamura et al. 2008, 2010, Nakamura \& Meier 2014, Cohen et al. 2015). However, there is a strong argument that supports them to be particle dominated at large distances to the formation region: The acceleration of the jet flow to relativistic speeds within the inner parsec can only happen if magnetic energy is converted into kinetic energy of the particles (see, e.g., Komissarov 2012), bringing the energy distribution close to equipartition. In addition, reverse shocks are quenched in strongly magnetised flows (see Mimica et al. 2009 for the case of GRB jets), which places serious doubts on the possibility that FRII jets, which show a reverse shock (hotspot), are magnetically dominated. This discussion is, however, out of the scope of this contribution. 
      
    Independently of the nature of the flow and the corresponding growing instability, the final outcome is the conversion of kinetic or magnetic energy into internal energy and the corresponding deceleration of the flow. In addition, in the non-linear regime, i.e., when the amplitude of the instability is of the order of the background values, mixing with the ambient medium can be enhanced by the deformation of the jet boundary. The resulting entrainment of slower and denser gas from the ambient medium produces efficient deceleration of the jet (see, e.g., Perucho et al. 2005, 2010). 
    
    Jets are surrounded by a dynamic environment when crossing the host galaxy, either their own backflow or the ISM. The perturbations that can be triggered along the jet are advected with the flow and, if coupled to an unstable KH or CD mode, grow in amplitude with distance. On the one hand, long wavelength perturbations can produce significant perturbations of the jet surface and produce strong shocks and entrainment, as has been proposed for the jet in the quasar S5~0836+710 (Perucho et al. 2012). On the other hand, short wavelength modes do not produce large distortions, but can enhance small scale turbulence and mixing at the jet boundary layer, which can progressively grow towards the jet axis (see, e.g., Laing and Bridle 2002, Perucho et al. 2005, 2010, Wang et al. 2009).

\subsection{Mass-load} \label{ml}
    
       Jets evolve in environments that are rich in gas clouds and stars within the first kiloparsec (e.g., the broad line and narrow line regions). These objects orbit the AGN and a number of them can impact and penetrate the jet: On the one hand, Araudo et al. (2010) estimated that clouds will enter into the jet if the shock crossing time through the cloud is longer than the penetration time of the cloud into the jet, which is  easier to happen in the case of denser clouds penetrating low power jets. On the other hand, stellar winds, even if weak, can equilibrate the jet ram pressure at a distance from the star (Komissarov 1994). Both clouds and stars ensure mass-loading of jets through the formation of bow-shocks and cometary tails that can be disrupted farther downstream (Bosch-Ramon et al. 2012, Perucho 2014, Perucho et al. in preparation). This scenario has been also claimed to explain observations at different frequencies in the jet of Centaurus~A (e.g., Worrall et al. 2008, Goodger et al. 2010, Wykes et al. 2013, 2015, M\"uller et al. 2014).

\begin{figure*}[!t]
\includegraphics[clip,angle=0,width=0.48\textwidth]{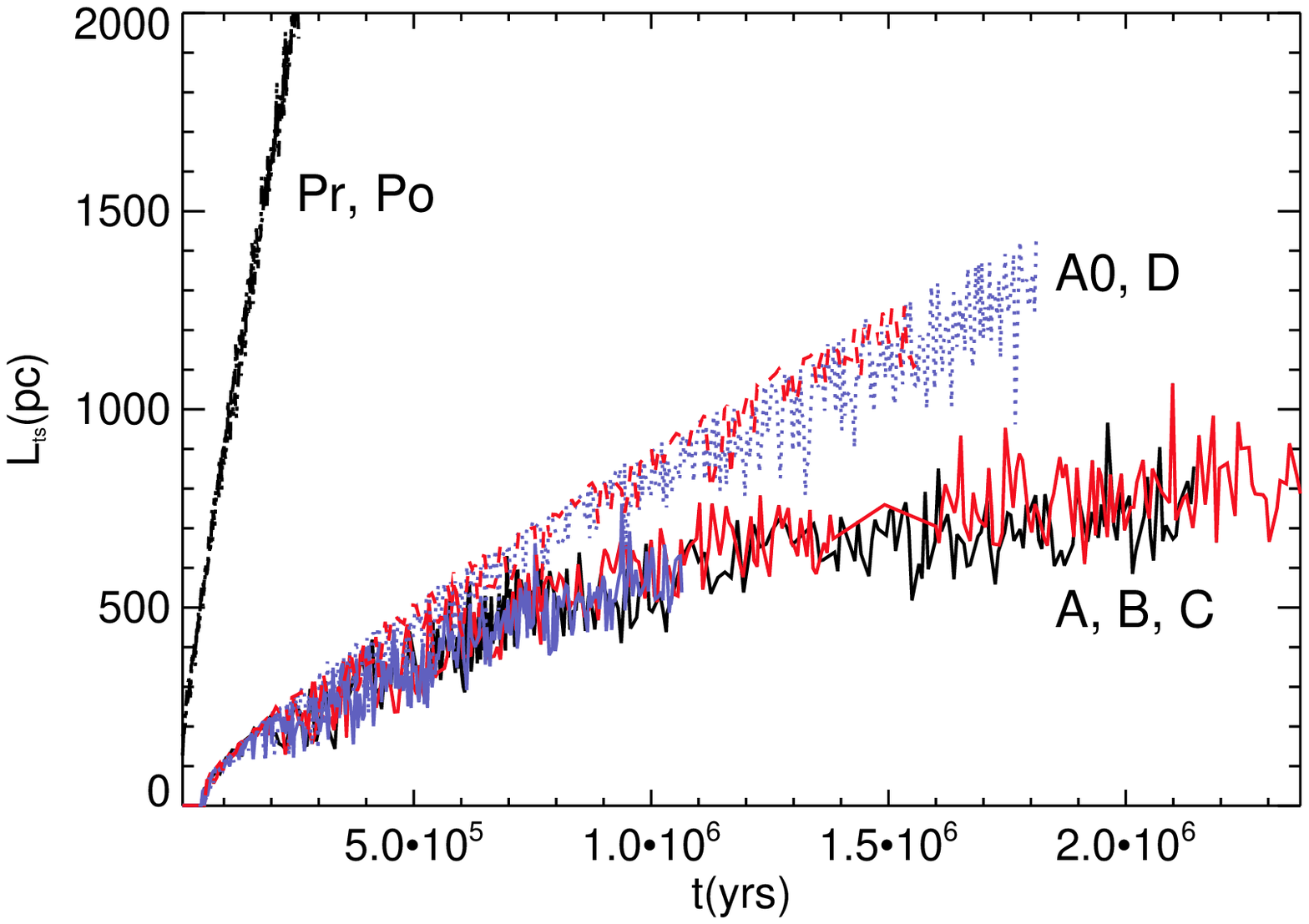} 
\includegraphics[clip,angle=0,width=0.48\textwidth]{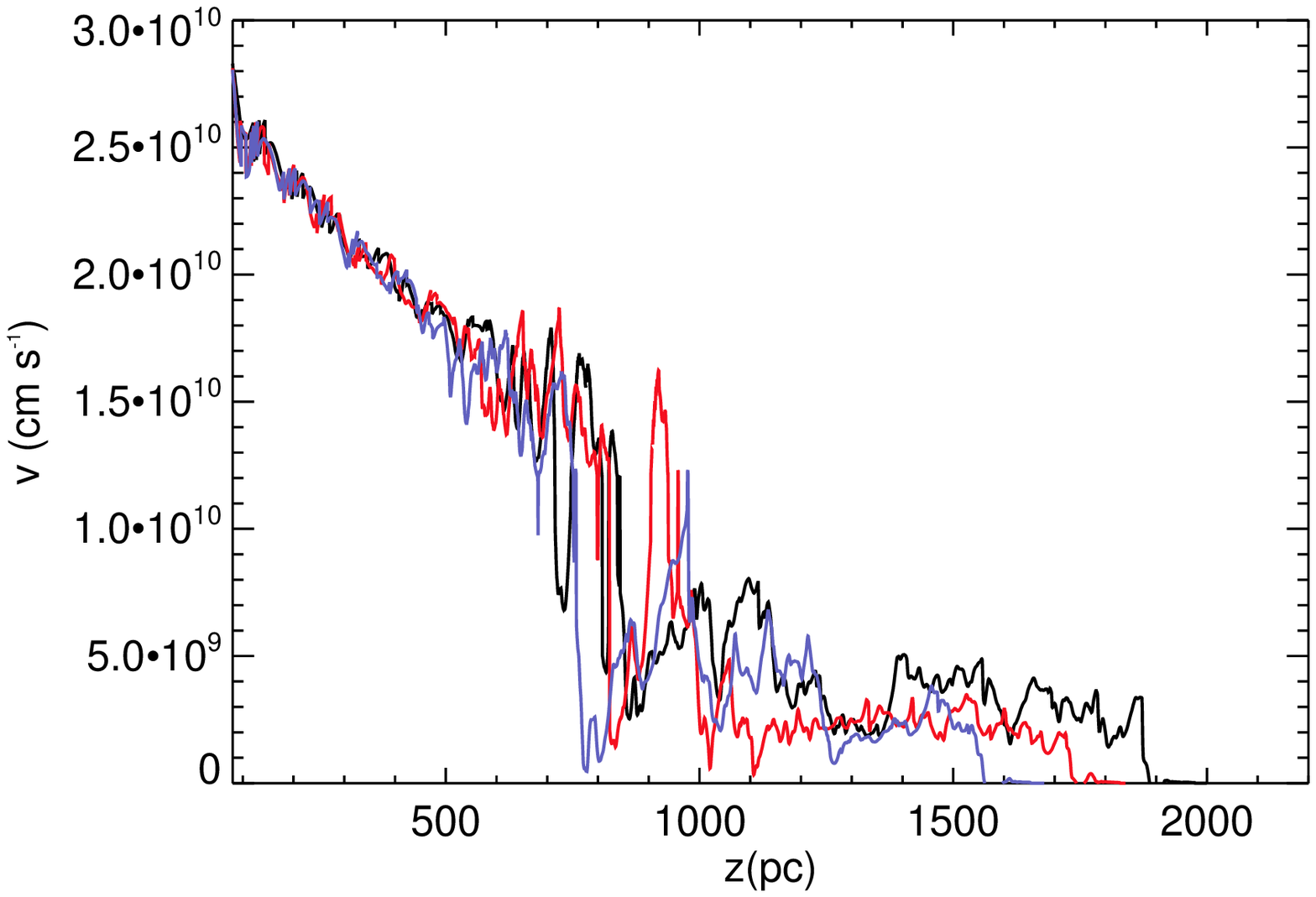} 
\caption{Left panel: Advance velocity of the head of the jet through the first stages of evolution for different jet powers; the dotted and dashed black lines represent the advance of powerful FRI jets (Pr, Po), the red-dotted and blue-dashed lines show the evolution of low-power jets with reduced or no mass-load (A0,D), and the solid lines represent the advance of the head with the same power than the latter, but with the mass-load expected from low mass, old stars (A, B, C, see the text for more details). Right panel: mean jet velocity versus distance for the low-power, mass-loaded jets, corresponding to A, B and C in the left panel.}
\label{fig1}
\end{figure*}

Komissarov (1994) studied the influence of mass loading by stellar winds in FR\,I jets and showed that this scenario can be treated as a hydrodynamical problem. Thus, this problem can be reduced to that of a distributed source
of mass that is injected into and thereafter advected with the jet
flow. Following this result, Bowman et al. (1996) solved the hydrodynamics equations for the case of a jet
in steady state, including a source term accounting for mass loading by a single type (typical of giant elliptic galaxies) stellar population. In particular, they focused on light and hot electron/proton jets and concluded 
that the studied cases could be efficiently decelerated by stellar winds. Finally, Hubbard \& Blackman (2006) presented an analytical study of the different ranges of stellar mass-loss rates and jet powers that could cause jet deceleration. In this work, the authors pointed out that a jet is efficiently decelerated when the kinetic energy needed to accelerate the mass loaded from stellar winds $\gamma_{\rm j} \dot{M}c^2$ per unit time (where $\gamma_{\rm j}$ is the jet Lorentz factor, and $\dot{M}$ is the mass-loading rate) is of the order of the jet kinetic luminosity, $L_{\rm j}$. If we assume a constant mass-loading rate per unit volume $q$ and a jet of constant radius $R_{\rm j}$, the characteristic length for jet deceleration is
\begin{eqnarray}
l_{\rm d} \simeq 10^3 \times\, \frac{1}{\gamma_{\rm j}} \times \, \! \! \left(\frac{L_{\rm j}}{10^{44} {\rm erg\,s}^{-1}}\right) \times \nonumber 
\\
\times \,\!\! \left(\frac{q}{10^{22} {\rm g\,yr}^{-1}{\rm pc}^{-3}}\right)^{-1} \times \, \!\!
\! \left(\frac{R_{\rm j}}{10 {\rm pc}}\right)^{-2} \!\! {\rm kpc}.
\label{eq:ld}
\end{eqnarray}
This expression can be only taken as indicative because part of the kinetic energy is used in heating of the loaded gas, and the Lorentz factor of the jet decreases as the jet is loaded along its path. In addition, the mass load rate changes (decreases) with distance to the nucleus, as the number of stars drops. Using the typical values for a single population of stars in giant elliptic galaxies (low mass, old stars) with low values of mass-loss ($10^{-11}\,{\rm M_\odot/yr}$) and no changes with distance, $l_d$ is of the order of hundreds of kiloparsecs, unless $L_j\,\ll \,10^{44}\, {\rm erg\,s}^{-1}$ or the stellar population includes a relevant number of massive stars with powerful winds. 

   Perucho et al. (2014a) presented a series of simulations following the calculations by Bowman et al. (1996) and showed that jets with powers $L_j\,\leq\,10^{42}\, {\rm erg\,s}^{-1}$ can be efficiently decelerated within the inner kiloparsec, which results in the jet head velocity asymptotically tending to zero. The left panel of Figure~1 shows the advance velocity of the head of the jet through the first stages of evolution for different jet powers: The dotted and dashed black lines represent the advance of powerful FRI jets ($L_j\simeq 10^{44}~\rm{erg/s}$), the red-dotted and blue-dashed lines show the evolution of low-power jets ($L_j\simeq 10^{41.5}~\rm{erg/s}$) with reduced or no mass-load, and the solid lines represent the advance of the head with the same power than the latter, but with the mass-load expected from low mass, old stars. The lines corresponding to the low-power, mass-loaded jets tend to zero velocity with time, and this implies that jets with this power are probably trapped within their host galaxies, though emitting in radio as long as the source is active. The right panel shows the mean jet velocity versus distance for these same jets: As the jets mass-load and decelerate, they expand, favoring the inclusion of more stars and further deceleration, until they become transonic and prone to the rapid growth of instabilities. The deceleration of the jet flow is continuous until the reverse shock, where it becomes subsonic and disrupts.

\subsection{Shocks}

    In this section I discuss the development of reconfinement shocks (see, e.g., Daly \& Marscher 1988, Falle 1991, Nalewajko \& Sikora 2009) and their effect. Jets evolve, during the first stages of their evolution, embedded within their own backflow (cocoon). This gas is typically hot and denser than the jet flow, so it is probably confining the jet via mild reconfinement cross shocks. Nevertheless, when the jet head leaves the density core, expands faster and its pressure drops. Taking into account that the sound speed in the cocoon is high, the pressure changes affect the whole region in relatively short times. Then, the jet can become over-pressured and expand. Given that the flow is supersonic, the expansion stops when it becomes under-pressured with respect to the ambient medium, and the information propagates towards the axis as a shock wave. The jump in the flow magnitudes depends on the original pressure difference, being larger for larger jet overpressure. An efficient conversion of kinetic energy into internal energy takes place at this shock. The flow becomes over-pressured again and expands adiabatically downstream, converting again internal energy into kinetic energy. If the overpressure is large enough a Mach disk can form on the jet axis. In this case, the conversion of kinetic energy into internal energy is so efficient that the flow becomes subsonic. 
    
\begin{figure*}[!t]
 \begin{center}
\includegraphics[clip,angle=0,width=0.8\textwidth]{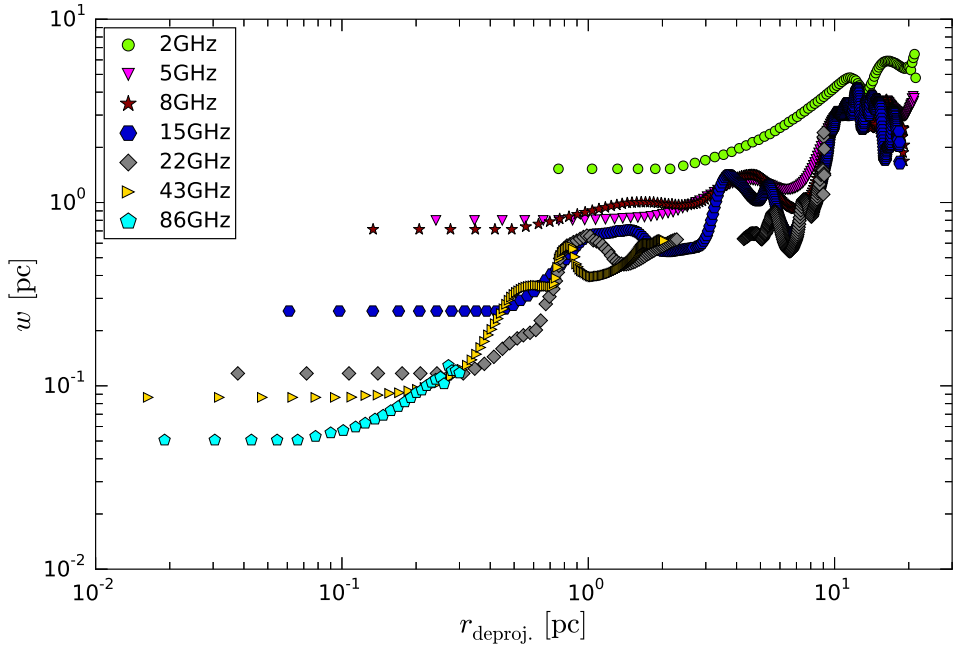} 
\caption{Jet width versus deprojected distance as measured at different frequencies for CTA~102 using very long baseline interferometry (Fromm et al. 2013b). The initial horizontal lines indicate the region where the jet is not resolved at a given frequency. Once the jet is resolved it shows the same radial size at all frequencies (0.3-1 mas in $r_{deproj}$). Beyond $r_{deproj}\simeq 1~\rm{mas}$ the jet width shows a plateau in a region in which it is well resolved. This is interpreted by the authors as a clear signature of jet recollimation. A new process of expansion and recollimation seems to follow farther downstream, as expected until the jet reaches equilibrium with the ambient medium.}
\label{fig2}
\end{center}
\end{figure*}

     From an observational perspective, Fromm et al. (2013a,b, see Fig.~\ref{fig2}) found evidence of the presence of reconfinement shocks at the parsec-scales in the blazar CTA~102, and Agudo et al. (2012) identified a stationary bright feature in the jet of 3C~120 as a recollimation shock, too.  At larger scales, Godfrey et al. (2012) identified periodic structures at the Megaparsec-scale jet of PKS 0637-752 as possible recollimation shocks. 

   Perucho \& Mart\'{\i} (2007) showed that the development of a Mach disk at kiloparsec scales could trigger the development of high amplitude oscillations and the eventual disruption and deceleration of the jet flow due to mixing. The development of such a disk depends on the opening angle of the jet, which is smaller for faster and less overpressured jets (Falle 1991, Fromm 2015). Although a strong poloidal magnetic field in the jet would prevent strong shocks to develop (e.g., Roca-Sogorb et al. 2009), it can be shown that Mach disks can develop in magnetised jets (Mart\'{\i} 2015), depending on the magnetisation of the ambient medium. It is important to stress that the size of the core of the galactic density profile, and the slope of this profile beyond the core play a fundamental role in this respect: Larger cores and/or shallow slopes of the King profile will not allow significant jet expansions (and contractions) of the jet, but steep slopes following a moderate core size ($r_{\rm c}\leq 1\,{\rm kpc}$, with $r_c$ the core size) can be the cause of eventual jet disruption before the jet reaches the large scales.  
       
\section{Evolutionary tracks} \label{et} 

   In this section I use the concepts exposed before to discuss possible evolutionary tracks of the different jet morphology types within the $P-D$ diagram. The jet injected power is invested into advance and expansion, and is hosted in the form of internal and kinetic energy kept by the jet particles and transferred to the ambient medium particles via shocks or mixing, or radiated away. Both the internal energy budget of the particles and that invested into acceleration of particles to relativistic energies are fed by the kinetic energy of the jet via transformation at waves, shocks and mixing. In particular, when the ambient medium is denser, the radio source expands more slowly and the pressure (both gas and magnetic) of the shocked region is higher at a given time than in the case of a faster expansion through a lower density medium. Therefore, it is important to stress that a strong jet-ambient medium interaction can be associated to a more efficient conversion of jet flow kinetic energy into radiated power owing to the work that the jet has to do to open its way through this medium. Scattering of the position of radio sources in the diagram is precisely expected due to possible differences in the ambient medium: In this way, a radio source with a higher kinetic power propagating through a low density medium could have lower values of $P$ than a radio source with a slightly smaller value of the kinetic power propagating through a denser medium. 

\subsection{FRI sources}
  
    FRI jet kinetic luminosities are typically $L_j\,\leq\,10^{44}\, {\rm erg\,s}^{-1}$. The parsec-scale jets of blazars, the proposed corresponding type-I AGN population to FRI radio-galaxies show superluminal velocities (e.g., Lister et al. 2009, Fromm et al. 2013a). There must thus be a process of deceleration of the jet flow from parsec to kiloparsec scales, as stated by the current paradigm for FR\,I jets (Bicknell 1984, De Young 1986, Laing 1993, 1996). From the modelling of observations of FRI jets, the most favoured options in the case of jets with powers $L_j\,\leq\,10^{43-44}\, {\rm erg\,s}^{-1}$ are either mass load via the development of small wavelength instability modes at the jet boundary with a mixing region that grows towards the jet axis, or a strong reconfinement shock at the flaring region (see, e.g., Laing \& Bridle 2002, Laing \& Bridle 2014). 

     During the first stages of evolution (when the size of the radio source size is of the order of 1~kpc) the morphology may be similar to that expected for a CSO (Perucho \& Mart\'{\i} 2007). When the bow-shock propagates far beyond this region, the jet deceleration due to the development of a recollimation shock or mass-load favours the growth of instability modes and the jet hot-spot is substituted by a terminal shock, detached from the bow-shock, after which the flow becomes turbulent (Perucho \& Mart\'{\i} 2007, Perucho et al. 2014a). Simulations show that this head is not steady, but undergoes periodical expansions and contractions, even if the injection rate is kept constant (Perucho \& Mart\'{\i} 2007)\footnote{This effect was also observed in non-relativistic MHD simulations (Lind et al. 1989)}. The bow-shock continues to propagate, driven by the overpressure generated by the injected material (see Perucho \& Mart\'{\i} 2007). Recent observational work on FRI jets (Kraft et al. 2007, Croston et al. 2007, Guidetti et al. 2011) also show that low-power, young radio sources are actually surrounded by bow-shocks. 
          
      The aforementioned result implies that not all compact symmetric radio sources will end up generating a large scale FRII source. It is also relevant to point out that the final measured size of the radio sources can be the size of the jet, if observed at GHz frequencies, and that the decollimated structure may not be observed at the same frequencies. This may lead to an overestimate of the number of compact (also symmetric) sources. Regarding the luminosity evolution, although there are so far no emission calculations based on simulations, the strong interaction with the ISM within the first kiloparsec should imply high luminosity. After the jet disruption, the total luminosity should decrease because the collimated jet does not show a strong interaction with the ambient medium as the bow-shock advances.
   
\subsubsection{A possible explanation for FR0 sources}

     FR0 sources have been recently introduced as a new class of radio sources (Baldi et al. 2015). They are characterised by being low-power and compact, showing no extended structure. I have discussed in Section~\ref{ml} that low power jets could be efficiently decelerated within the inner two kiloparsecs in the host galaxy. These low-power jets can be injected with relativistic or mildly relativistic velocities, but being very dilute (in order to account for the low power), dominated by pairs and/or magnetic field. The evolution of these radio sources in the $P-D$ diagram would be qualitatively similar to that given above for FRI sources, but shifted down in luminosity and falling in $P$ at lower values of $D$, as compared to FRI's. Thus, one could speculate with the possibility that the number of low-power radio sources is larger than that of high-power radio sources. If, in addition, low-power sources usually grow to smaller values of $D$ (in terms of the physical processes described in the previous section) before they get exhausted towards the end of their activity period, one could at least qualitatively explain the excess of compact sources (e.g., O'Dea \& Baum 1997, Alexander 2000).

\begin{figure*}[!t]
\includegraphics[clip,angle=0,width=\textwidth]{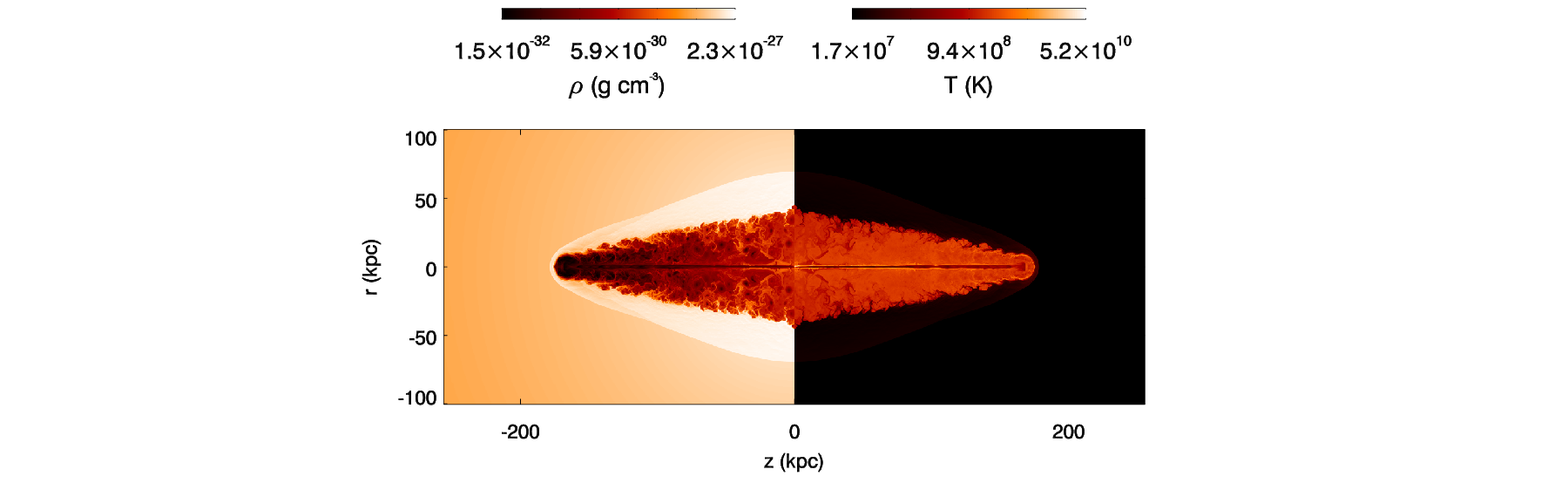} 
\includegraphics[clip,angle=0,width=\textwidth]{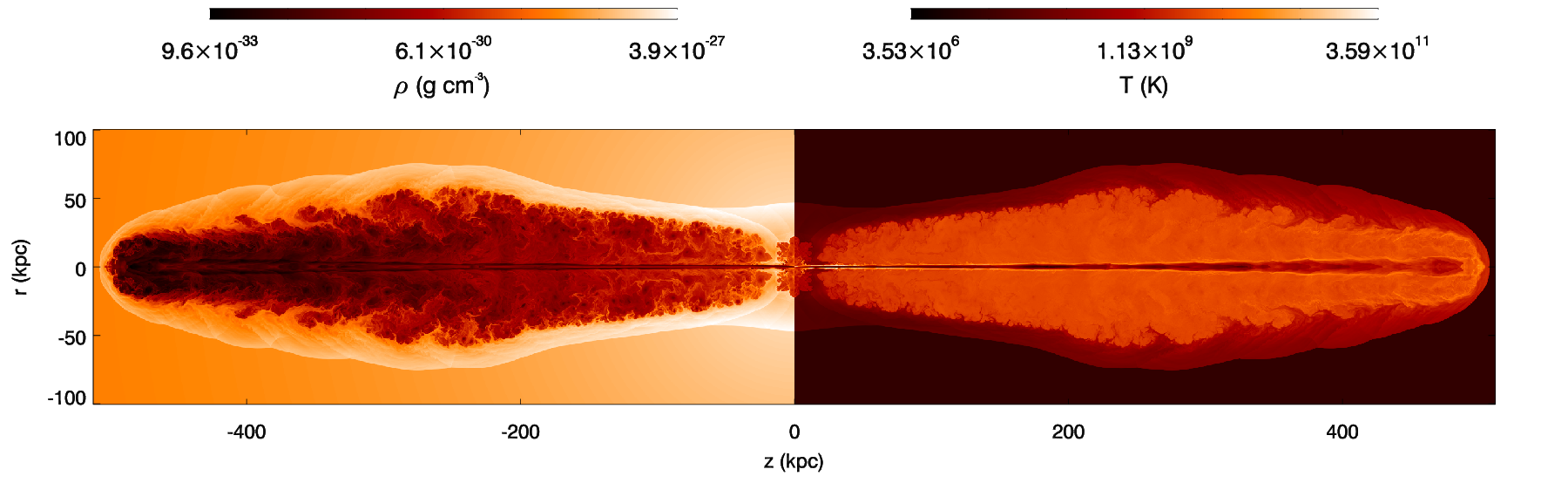} 
\caption{Rest mass density (left) and temperature (right) for a snapshot of two-dimensional axisymmetric simulations of long-term evolution through a King-profile in density and pressure. The upper panel shows a leptonic jet with kinetic power $L_j\simeq 10^{44}~\rm{erg/s}$ at $t\simeq 50~\rm{Myr}$, whereas the lower panel shows a leptonic jet with kinetic power $L_j\simeq 10^{46}~\rm{erg/s}$ at $t\simeq 16~\rm{Myr}$. From Perucho et al. (2014b).}
\label{fig3}
\end{figure*}

\subsection{FRII sources}

    FRII sources show collimated jets through hundreds of kiloparsecs. The strong interaction between the jet and the environment, as revealed by the presence of a hot-spot, implies that a non-negligible portion of the jet energy is converted into gas internal energy and radiative power. Taking into account that these jets have large powers ($L_j > 10^{44}\, {\rm erg\,s}^{-1}$), these sources should always be found at the top of the $P-D$ diagram distribution. The exact position will also be determined by the gas distribution of the ISM and IGM through which it propagates. In this respect, a decrease in luminosity is expected at large scales due to the decrease in density of the ambient medium gas. The same reason explains the observed jet head acceleration between 1 and 10 kpc in numerical simulations (Perucho et al. 2014b), which has also been claimed by Kawakatu et al. (2008) to explain the observed evolution of the cocoon sizes for radio sources before and after this linear size. Wagner et al. (2012) have shown that powerful jets can have a strong impact on a highly inhomogeneous medium within the first kiloparsec, generating very complex structures and also making the jet advance slower until the jet head manages to drill through the whole region.


   RHD simulations of larger scales, studying the evolution of jets once they have crossed the inner kpc, show that a large fraction of the injected energy in a jet is transferred to the ambient medium (Perucho et al. 2011, 2014b) via a strong shock and mixing at the contact discontinuity between the shocked jet gas and shocked ambient medium. Figure~\ref{fig3} shows a snapshot of two-dimensional, axisymmetric simulations of two jets with different power. It displays feast-mass density and temperature. The jet generates an under-dense and very hot region (cocoon) with a size that depends on the power, as well as the shape of the cocoon and bow shock. In the case of two-dimensional simulations, the shock heating dominates the process of energy injection in the ambient medium, but this should be confirmed by three-dimensional simulations, in which mixing may be enhanced by the growth of instabilities (Perucho et al., in preparation).

\section{Discussion} \label{di}
  
  Numerical simulations show that the evolution through the first kiloparsec is very different to that beyond this region (Wagner et al. 2012), in agreement with an argument given in the nineties (e.g., O'Dea \& Baum 1997). In this region, the jet propagates through a denser and more inhomogeneous ambient medium and the injected power is invested primarily in shocking and heating the gas that is displaced. A part of this energy must be radiated away, with the exact amount depending on the efficiency of the process, still unknown. Thus, the radio sources are brighter when they cross this inner region, and can even increase their brightness as the amount of injected energy grows faster than the volume of the evacuated region, so that the pressure within the region increases. 
  
   A rule-of-thumb criterion to know the fate of radio sources was given by Kawakatu et al. (2009). The authors calculated the distance at which a jet would become transonic due to the work done on the ambient medium and obtained the critical ratio $L_j/\bar{n}_a$ (where $\bar{n}_a$ is the number density at the location where the slope of the density profile changes) at which the jet propagation would become subsonic and a critical value over which the jet head would be supersonic at all scales ($L_j/\bar{n}_a\simeq 10^{45}\,\rm{erg/s \, cm^3}$, see the reference for details). Numerical simulations show that this represents a good first-order guess to the fate of an extragalactic jet. However, there are nonlinear effects that can change this picture during the first kiloparsecs, where the numerical simulations show that the deceleration distances can be different to those predicted on the pure basis of the jet power and density, if processes such as mass-load by stellar winds are taken into account (Perucho et al. 2014a). 
   
  Physical processes like the growth of instabilities, mass-load by the local stellar population, or the development of shocks can completely change the development of jets farther away and also in the $P-D$ plane. These processes result in a clear trend that forbids low-power radio sources to propagate beyond a few kiloparsecs, with the exact distance depending on the jet power and properties of the host galaxy. The trend can be defined by the following facts: 1) Unstable modes develop faster in jets with low Mach numbers (Hardee 1987, Hardee et al. 1998, Perucho et al. 2005), more likely to be the case for low-power jets; 2) the deceleration by mass-load is more efficient in the case of low-power jets (Perucho et al. 2014a), and 3) the development of strong recollimation shocks depends on jet overpressure, but also on the jet opening and reconfinement angles, which also depend on the jet Lorentz factor (the largest the Lorentz factor, the smaller the opening angle, see, e.g., Falle 1991, Fromm 2015), and therefore, jets with smaller values of the Lorentz factor are more prone to develop strong shocks. From the latter three items we can state that low power jets are probably related to the radio sources that populate the mid-bottom of the $P-D$ diagram. On the contrary, high-power sources must populate the top of the diagram at all scales. The transition from one region of the map to the other must be populated by in-betweeners like hybrid radio sources (HYMORS, Gopal-Krishna \& Wiita 2000) or the powerful quasar S5~0836+710, which has been reported to be possibly disrupted by the growth of a helical KH mode (Perucho et al. 2012). 
   
   Only when the source of the jet is extinguished or the injected power decreases, the radio-luminosity drops dramatically, which is common to all jet powers and sizes. Taking into account that the jet flow propagates at a large fraction of the speed of light, even if the head of the jet is at several hundred kpc, the time needed for the information of the decrease in power to reach the head is $\simeq 1\,\rm{Myr}$, which is much less than its typical lifetime at these scales $~10-100\,\rm{Myr}$. The moment in which the information reaches the head could represent the start of an eventual transition from FRII to FRI morphology, if the activity is kept longer in time, albeit with lower power than during the FRII phase. Nevertheless, the claim of FRII sources being the progenitors of large-scale FRI sources (Wang et al. 2008) seems to be placed in doubt under the light of intrinsic differences in the properties of the host galaxies (Sadler et al. 2014). In this case all radio sources would turn into relics at the end of their active cycles. 
   
     A re-activation of the source (e.g., Bogda\'n et al. 2014) can lead to a very different situation that still needs to be explored via numerical simulations of relativistic outflows with duty cycles (e.g., Walg et al. 2014). For instance, a compact radio source with lower luminosity than the original jet could be generated, even if it has the same power, if it propagates through a lower density medium after the first active phase has evacuated the region. It should be stressed that the gas reservoir around the central engine is to play a crucial role in the jet-power evolution along the history of a given radio-source: short-term changes up to short jet generation event could be provided by the fall of molecular clouds or minor mergers onto the central black hole, and long-term changes can be given by the lack of inflowing gas if no merging processes take place at the host galaxy.       
   
   Summarising, the jet power and its possible evolution seem to rule the long-term evolution of radio sources and their paths within the $P-D$ diagram. However, at the boundary between low-power and high-power jets ($10^{43}-10^{45}\,\rm{erg/s}$) other factors such as the properties of the ambient medium, the growth of unstable KH (or CD) modes or even the stellar population of the host galaxy can play a fundamental role in the evolution in size and luminosity of the radio sources. A simple prediction of the idea exposed here is that low-power jets are more common than high-power jets, which would directly explain the excess of compact sources in the number counts, as low-power jets would tend to be more compact under the explanation given here. Alexander (2000) gave the first hint in this direction, suggesting that frustrated, compact sources are short lived sources. It is possible, however, that some of those sources are not short-lived, but relatively old and propagating very slowly. 
   
   3D simulations of relativistic (magnetised) jets in realistic environments should be envisaged in order to confirm the description given in this contribution. We run now large-scale, 3D simulations of jet evolution from 1 kpc to several hundred kpc through a King-profile ambient medium and including small scale inhomogeneities with the aim of characterising the propagation of the jet and its effect on the environment. Although the jets are perturbed at injection, their large power and inertia given by a moderately high Lorentz factor allow collimation at a distance $\simeq$100~kpc from injection, in agreement with our previous statements (Perucho et al., in preparation). 
   
   Further work should include a combination of multi-wavelength studies of the host galaxies with an improvement of the setups of the numerical simulations in order to provide a better physical description of these systems. The aim, from the numerical perspective, should be to produce results that can be compared with the state-of-art observations that try to relate the propagation of jets at these scales with the feedback (positive or negative) on the host galaxy (e.g., O'Dea 1998, Labiano 2008, Labiano et al. 2008, Ogle et al. 2010, Willett et al. 2010).   


\acknowledgements 
MP is a member of the work team of the Spanish MI\-NE\-CO grants AYA2013-40979-P and AYA2013-48226-C3-2-P. I acknowledge the SOC and LOC of the 5th Workshop on CSS and GPS radio sources. I acknowledge Jos\'e-Mar\'{\i}a Mart\'{\i} and the anonymous referee for useful comments to the manuscript. The research leading to these results has received funding from the  European Commission Seventh Framework Programme (FP/2007-2013) under grant agreement No 283393 (RadioNet3). 

%
%

\end{document}